%% file: vince-cr.tex
\documentclass{article}
\usepackage{ijcai22}

\usepackage{times}

\usepackage{soul}
\usepackage{url}
\usepackage[hidelinks]{hyperref}
\usepackage[utf8]{inputenc}
\usepackage[small]{caption}
\usepackage{graphicx}
\usepackage{amsmath}
\usepackage{booktabs}
\usepackage{threeparttable}
\usepackage{multirow}
\usepackage{amssymb}
\usepackage{bbding}
\usepackage{wasysym}
\newcommand{\comment}[1]{}

\newcommand{\tabincell}[2]{\begin{tabular}{@{}#1@{}}#2\end{tabular}}

\title{Deep Learning Meets Software Engineering:\\
A Survey on Pre-Trained Models of Source Code}

\author{
Changan Niu$^1$\and 
Chuanyi Li$^1$\and
Bin Luo$^1$ and
Vincent Ng$^2$\\
\affiliations
$^1$State Key Laboratory for Novel Software Technology, Nanjing University, Nanjing, China\\
$^2$Human Language Technology Research Institute, University of Texas at Dallas, Richardson, Texas, USA\\
\emails
niu.ca@outlook.com, \{lcy,luobin\}@nju.edu.cn, vince@hlt.utdallas.edu
}

\begin{document}

\maketitle

\begin{abstract}
Recent years have seen the successful application of deep learning to software engineering (SE). In particular, the development and use
of pre-trained models of source code has enabled state-of-the-art results to be achieved on a wide variety of SE tasks. 
This paper provides an overview of this rapidly advancing field of research and reflects on future research directions. 
\end{abstract}

\section{Introduction}
\label{section:introduction}

Once upon a time the state of software intelligence in software engineering (SE) was very rudimentary, with many of the decisions supported by gut feeling and at best through consultation with senior developers \cite{hassan2010software}. As a wealth of data has been generated in the software development and evolution lifecycle over the years, the software development and evolution paradigm has also shifted from human experience-based to data-driven decision making. While AI researchers are fully aware of the impact deep learning has on AI application domains such as computer vision and natural language processing (NLP), many are not aware of the extensive and successful applications of deep learning technologies to SE tasks in recent years.

Though successful, the application of deep learning is not without challenges. One such challenge concerns the need for a large, typically costly-to-obtain, annotated training set to train the millions or even billions of parameters in deep neural networks. To address this data annotation bottleneck, NLP researchers have come up with an idea that can arguably be considered a breakthrough in recent deep learning research, namely {\em pre-training} \cite{dai:nips15,howard:acl18,peters2018deep}. Rather than training a model from scratch (i.e., with randomly initialized network weights), which typically requires a lot of task-specific annotated data, one can first {\em pre-train} it on one or more so-called {\em self-supervised} tasks (i.e., tasks for which annotated data can be automatically generated and therefore large amounts of training data are readily available) so that its weights encode general linguistic and commonsense knowledge about language, and then the resulting {\em pre-trained} model can be {\em fine-tuned} to learn the target task using (a potentially small amount of) task-specific annotated training data in the usual supervised manner. A large number of pre-trained language models have been developed and widely used in NLP, such as BERT \cite{devlin2019bert}, XLNet \cite{yang2019xlnet}, RoBERTa \cite{liu2019roberta}, ELECTRA \cite{clark2019electra}, GPT-2 \cite{gpt2}, T5 \cite{raffel2020t5}, and BART \cite{lewis2020bart}.

Can these pre-trained models be applied to SE tasks? Since source code can be viewed as a sequence of code tokens in the same way that natural language (NL) can be viewed as a sequence of word tokens, we can in principle retrain these models on source code and apply them to SE tasks. In practice, this is not ideal, as there are code-specific characteristics that may not be properly taken into account by these models.
For instance, source code is not as homogeneous as NL: it is composed of both the code in a function body, which is written in programming language (PL), as well as optional comments written in NL. Treating both code and comments in a uniform manner (i.e., as a sequence of tokens) may not be the best way to exploit the two sources of information. In addition, code has syntactic structures (as defined in Abstract Syntax Trees (ASTs)) and semantic structures (as defined in Control Flow Graphs (CFGs)). While a few syntax-aware pre-trained models are recently developed in the NLP community (e.g., Xu et al.~\shortcite{xu2021syntax}), the majority of existing pre-trained models fail to exploit structured information. Consequently, SE researchers have developed a number of pre-trained models of source code (CodePTMs) that take into account the characteristics specific to source code in the past few years.

Our goal in this paper is to raise the awareness of the AI audience on the impact that AI technologies --- in this case the development and use of pre-trained models --- have on SE, an important AI application domain, specifically by providing them with a survey of the recent development of CodePTMs and their successful application to SE tasks. We believe this survey will be of particular interest to (1) NLP researchers, especially those focusing on text summarization and generation, since many SE tasks (e.g., code summarization) involve NL generation; and (2) applied machine learning researchers, since the development of these models could have a big impact on SE.
Though our target audience is AI researchers, we believe this paper could also be of high interest for the SE technology providers, raising their awareness on the added value AI technology could have in augmenting SE tooling to leverage the increasing complexity of software systems.

\input{table_tasks}

\section{SE Tasks, Datasets, and Evaluation Metrics}
\label{section:se_tasks}

SE studies problems concerning the design, development, maintenance, testing, and evolution of software systems. Table~\ref{table:application} enumerates the key SE tasks to which pre-trained models have been applied. As can be seen in the first two columns, we classify each task along two dimensions: (1) whether the task concerns {\em understanding} ({\bf Und.}) or {\em generation} ({\bf Gen.}); and (2) the type of input assumed by the task and the type of output produced ({\bf I-O}), where {\bf C}, {\bf NL}, and {\bf V} denote code, natural language, and extracted/predicted value, respectively.

In addition, Table~\ref{table:application} shows for each task the benchmark dataset(s) and the corresponding evaluation metric(s). These metrics are fairly standard. For retrieval and classification tasks, metrics such as Acc (Accuracy~\cite{cubert}), Acc@k (Accuracy computed over the top $k$ predicted answers~\cite{watson2020learning}), Precision(P)/Recall(R)/F1~\cite{nafi2019clcdsa}, MRR (Mean Reciprocal Rank~\cite{husain2019codesearchnet}), MAP@R (Mean Average Precision~\cite{mou2016convolutional}), and NDCG (Normalized Discounted Cumulative Gain~\cite{nafi2019clcdsa}) are typically used. For generation tasks, metrics developed in the NLP community for summarization and translation tasks, such as BLEU~\cite{papineni2002bleu}, ROUGE-L (RL)~\cite{haque2020improved}, METEOR~\cite{hu2018summarizing}, CIDER~\cite{codedisen}, and EditSim~\cite{gpt-c} (an edit distance-based metric),
as well as variants developed in the SE community, such as CodeBLEU (CBLEU)~\cite{ren2020codebleu}, are used.

\section{CodePTMs}
\label{section:overview}

In this section, we provide an overview of 20 CodePTMs recently developed in the SE community. To enable the reader to better understand their similarities and differences, as well as their relative strengths and weaknesses, we classify them along four dimensions, as described below.

\subsection{Architecture}
\label{section:architecture}

\input{table_pre_training_task}

First, existing CodePTMs differ in terms of the underlying network architecture. 
To understand network architectures, we need to briefly introduce the concepts of encoding and decoding.
An encoder encodes an input sequence as a fixed-length vector representation, whereas a decoder generates an output sequence based on the representation of an input.

Rather than designing new network architectures, SE researchers base the design of CodePTMs on existing architectures. Broadly, these architectures can be divided into four categories: (1) {\bf Long Short-Term Memory} (LSTM~\cite{hochreiter1997long}), which is a classical recurrent neural network architecture, (2) {\bf Transformer} (TF) \cite{Vaswani2017attention}, which is a comparatively newer encoder-decoder architecture\footnote{Recall that an encoder-decoder architecture is commonly used for sequence-to-sequence tasks, where the encoder encodes an input sequence as a fixed-length, typically task-specific, representation, and the decoder then generates an output sequence token by token based on the input and the tokens that have been generated so far.} that is faster to train and can better capture long-distance dependencies than LSTM; (3) {\bf Transformer-Encoder} (TE), which corresponds to the architecture of the encoder part of TF; and (4) {\bf Transformer-Decoder} (TD), which corresponds to the architecture of the decoder part of TF. While it is possible to use encoder-only models (such as TE) and decoder-only models (such as TD) for sequence-to-sequence (seq2seq) tasks, it has been shown to be disadvantageous and impractical to do so~\cite{spt-code}. In particular, encoder-only models and decoder-only models are disadvantaged when applied to generation/decoding and classification tasks, respectively.

\subsection{Modality}
\label{section:modality}

When using a neural model to process source code, being able to integrate the NL embedded in the code (e.g., documentations, variable names) and the code structure (e.g., ASTs) can improve the model's ability to understand the code~\cite{ernst2017natural,hu2018summarizing,leclair2019neural,zugner2021language}. Therefore, the use of NL and code structure as inputs in addition to the code itself has become a common practice in CodePTMs. As Code, NL, and Structure differ in representation and processing, they can be viewed as features of different input modalities. Hence, along the second dimension, we divide CodePTMs into three categories --- unimodal ({\bf Uni}), bimodal ({\bf Bi}), and multimodal ({\bf Multi}) --- based on the number of input modalities they employ.

When a model employs more than one input modality, we can either (1) concatenate the features extracted from different modalities to form a single training instance or (2) use the features extracted from different modalities to create different training instances. We refer to these two strategies as {\em Together} and {\em Standalone}, respectively. As can be imagined, an advantage of {\em Together} over {\em Standalone} is that the former allows cross-modal representations to be learned by a model.  

\subsection{Pre-Training Tasks}
\label{section:pre_training_task}

Along the third dimension, we differentiate CodePTMs based on the tasks used to pre-train them. At a high level, we can divide these tasks into two categories depending on whether the task originates in NLP ({\bf NLP}) or is specifically designed for source code ({\bf SE}), as shown in Table~\ref{table:pretrainingtask}.

As can be seen from the table, the NLP pre-training tasks can be subdivided into four categories: (1) Language modeling ({\bf LM})~\cite{qiu2020pre}, which refers to the collection of tasks that aim to predict a given word given the surrounding context; (2) Masked Language Modeling ({\bf MLM})~\cite{devlin2019bert}, which refers to the collection of tasks that aim to predict the masked tokens; (3) Denoising Auto-Encoding ({\bf DAE})~\cite{lewis2020bart}, which aim to recover the original (i.e., uncorrupted) text from corrupted text; and (4) Contrastive Learning ({\bf CTL})~\cite{contracode}, which allows a model to learn which data points are similar or different. The SE pre-training tasks, on the other hand, can be subdivided into three categories according to their input modalities: (1) Code-Aware ({\bf CA}) tasks, which aim to mine latent information from code text; (2) Structure-Aware ({\bf SA}) tasks, which aim to learn representations of the code structure; and (3) Cross-Modal-Aware ({\bf CMA}) tasks, which seek to acquire knowledge from multiple input modalities. The CMA tasks can be further subdivided into three categories based on which input modalities are involved, namely Code-NL ({\bf CN}), Code-Structure ({\bf CS}) and Code-NL-Structure ({\bf CNS}).

\input{table_cptms_taxonomy}

When more than one task is used to pre-train a CodePTM, the tasks involved can be learned {\em simultaneously} (i.e., each data instance supports all of the tasks involved\footnote{A data instance {\em supports} a task if the task's loss can be computed based on the instance. For example, a code-only data instance (i.e., a code snippet without the paired docstring) supports both MLM and NSP because the losses of both tasks can be calculated based on the code snippet. However, it does not support BDG because the code-docstring alignment is needed by BDG.} and the task losses can be jointly minimized), {\em sequentially} (i.e., the model is first trained on the first task for a specified number of steps and then trained on the remaining tasks one by one), or {\em alternately} (i.e., 
the tasks are randomly optimized as batches of the data instances corresponding to a particular task are selected at random during training). Hence, simultaneous pre-training holds the strictest requirements on the data and the tasks because it requires that for each data instance, all the pre-training tasks can be completed in one forward propagation such that their losses can be added to form the final optimization objective and jointly minimized during backward propagation. In other words, if it can perform simultaneous pre-training, it will also be possible to perform sequential/alternate pre-training but not vice versa. Nevertheless, the selection of a pre-training strategy in existing CodePTMs seems random when multiple options are available\footnote{For example, IT in CodeT5 can be pre-trained simultaneously with any of the other tasks, but it is still pre-trained alternatively.}.

\subsection{Programming Languages}
\label{section:language}

Along the last dimension, we categorize CodePTMs depending on whether they are pre-trained on one PL
({\bf Monolingual (Mono)}) or multiple PLs ({\bf Multilingual (Multi)}). 

\input{table_cptms_details}

\subsection{Categorization and Pre-Training Details}

The first five columns of Table~\ref{table:taxonomy} categorize 20 CodePTMs along the four dimensions discussed in the previous subsections, namely Architecture ({\bf Arch.}), Modality ({\bf Mod.}, Pre-Training Tasks, and Programming Languages ({\bf PL}). We believe this categorization can help the reader better understand the similarities and differences between different CodePTMs.

Note, however, that Table~\ref{table:taxonomy} only provides a {\em high-level} categorization of the CodePTMs. For instance, we still do not know which two input modalities are used by a bimodal CodePTM, and neither do we know which PLs are used to pre-train a multilingual CodePTM. Table~\ref{table:codeptm_detail} fills this gap by providing the details of how each CodePTM is pre-trained. Specifically, {\bf CodePTM} cites the paper that proposed each CodePTM, whereas {\bf Input}, {\bf Objective}, and {\bf Dataset} show the input modalities, the pre-training tasks, and the PLs involved in pre-training each CodePTM. The datasets can be divided into four types, namely, {\em GitHub Repos} (a dataset obtained from GitHub, e.g., JS GitHub Repos is a dataset built by GitHub JavaScript repositories), {\em BigQuery} (a platform that includes activity from over 3M open source GitHub repositories, e.g., ``Python from BigQuery'' is the dataset collected by querying Python functions on BigQuery), {\em CodeSearchNet}~\cite{husain2019codesearchnet} (a dataset that is obtained by scraping open-source repositories and pairing individual functions with their docstrings and which includes more than 6.4M codes of 6 PLs including Java, Python, JavaScript, PHP, Go and Ruby), and {\em CLCDSA}~\cite{nafi2019clcdsa} (a dataset collected from Online Judge (OJ) sites across four PLs (i.e., Java, Python, C\# and C++) where functionally similar solutions written in different PLs are available for a given problem). 

\section{Discussion}
\label{section:application_analysis}

Next, we explore the relationship between CodePTMs (Section~3) and SE tasks (Section~2). The right half of Table~\ref{table:taxonomy} depicts this relationship by showing whether a CodePTM has been applied to a particular SE task, and if so, which benchmark dataset(s) it has been evaluated on and whether state-of-the-art (SOTA) results have been achieved. Below we discuss our key observations, which are based in part on Table~\ref{table:taxonomy} and in part on conclusions drawn from the literature. 

\vspace{-1mm}
\paragraph{Architecture.} As can be seen in Table~\ref{table:taxonomy}, TE-based CodePTMs are applied mostly to Understanding tasks, whereas TD- and TF-based CodePTMs are applied mostly to Generation tasks. This is understandable. As mentioned in Section~3.1, encoder-only models are disadvantaged when applied to Generation tasks. The reason is that they can only map an input sequence to an output sequence with a priori known length, but for Generation tasks the output length is typically not known a priori. In contrast, the presence of decoders in TD- and TF-based CodePTMs naturally makes them more suited to Generation tasks.

\vspace{-1mm}
\paragraph{Modality.} We make two modality-related observations. First, for CodePTMs that use structured information as input (e.g., features extracted from DFGs, ASTs, and AEI\footnote{AEI (Abstract Environment Information) describes a program's semantics with a mathematical characterization of its behaviors.}), removing such information from the input always reduces their performances on downstream SE tasks~\cite{graphcodebert,codedisen,syncobert,treebert}.

Second, the use of NL as an input modality appears to contribute positively to model performance on a downstream task only if NL is present in the input or output of the task~\cite{codebert,spt-code}. Otherwise, the use of NL could lead to a performance deterioration~\cite{cotext,spt-code}. For example, CodeT5, which is pre-trained using NL and Code, achieves SOTA results on all the NL-related SE tasks to which it is applied (e.g., TL, SU, and MN), but it is surpassed by SynCoBERT, which is pre-trained only on Code, in performance on CD, a Code-related-only task. 

\vspace{-1mm}
\paragraph{Pre-training tasks.} We make two pre-training tasks-related observations. First, after fine-tuning on task-specific training data, a pre-trained model generally yields better results on SE downstream tasks than its "no pre-training" counterpart that is trained only on task-specific training data, and the discrepancy in their performances is especially obvious when the amount of task-specific training data is small~\cite{zhou2021assessing,cubert,c-bert,dobf}. This is true even when the underlying pre-trained model is taken from the NLP domain, such as RoBERTa, without pre-training it again on source code~\cite{codebert,plbart}.

Second, keeping the pre-training task's type as similar as possible to that of the downstream task tends to yield the best results. Theoretically, pre-training will be beneficial for a downstream task precisely when the knowledge learned during pre-training can be successfully exploited when the model learns the downstream task. Such knowledge transfer tends to be more effective if the pre-training task is closer to the downstream task. For instance, for Understanding tasks, it is better to use a pre-training task that is also an Understanding task, such as MLM. Note that MLM is used by all the models that achieve SOTA results on Understanding tasks, such as CuBERT. In contrast, (I)MASS, which focuses on Generation, tends to work much better as a pre-training task than (I) MLM, which focuses on Understanding, on seq2seq downstream tasks such as code summarization~\cite{treebert}.

\vspace{-1mm}
\paragraph{Programming languages.} We make two PL-related observations. First, knowledge transfer tends to be a lot more effective if a CodePTM is trained on a PL that is syntactically similar to the one used in the downstream task. In contrast, knowledge learned by a CodePTM from PLs that are syntactically different from the one used in the downstream task may even lead to the performance degradation. For instance, PLBART, which is pre-trained on Java and Python, performs better on C\# code translation but worse on PHP code summarization than RoBERTa, a PTM that is trained on NL text only. The reason is that C\# is syntactically similar to Java, while PHP has a syntax mismatch with Java and Python.

Second, multilingual pre-training and fine-tuning generally yield better results than their monolingual counterparts. For example, CodeT5, which is pre-trained on 6 PLs, outperforms T5-learning and DeepDebug, both of which are only pre-trained on Java, on code translation. In addition, when performing multilingual pre-training, {\em language-aware} pre-training, where the training instances that belong to different PLs are being differentiated (by adding language-specific symbols to the input or appending a language-type embedding to each token, for instance), tend to yield a pre-trained model that can better discriminate between PLs than {\em language-agnostic} pre-training, as demonstrated via GPT-C on code completion.


\section{How Effective are CodePTMs?}

CodePTMs have been successfully applied to a variety of SE tasks, but {\em how} effective are they? To enable the reader to gain insights into this question, we present some quantitative results in this section. More specifically, we show in Table~\ref{table:sota} the best result achieved by a CodePTM on each commonly used evaluation dataset for each SE task (see the "Best CodePTM" column). To help the reader gauge the effectiveness of CodePTMs, we show in the "Best non-CodePTM" column the best result achieved by an approach that does not involve pre-training on each dataset. As can be seen, many of the best non-CodePTM-based approaches are neural models that involve Tree-LSTM and Transformer, for instance. The last column of the table  shows for each dataset the {\em relative error reduction rate}, which is computed as the error reduced by the best-performing CodePTM relative to the error made by the best non-CodePTM-based system on the dataset. A positive value indicates that the SOTA result is achieved using a CodePTM. 
As can be seen, the SOTA results on all of the datasets are achieved using CodePTMs, with the relative error reduction rates ranging from 0.9--78.7 when expressed in percentages. These results provide suggestive evidence that CodePTMs are a promising approach to a wide variety of SE tasks. Nevertheless, it is clear that CodePTMs are more effective at relative error reduction on certain SE tasks/datasets than other tasks/datasets. Additional analysis is needed to determine the reason.
\input{table_sota_new}

\section{Concluding Remarks}
\label{section:outlook}

Though CodePTMs have proven their success in SE, we believe that they have not reached their full potential. In this section, we outline some promising future directions.

\subsection{Thinking beyond NLP}
\label{section:outlook_outside_nlp}

\paragraph{Tokenization and embedding.}
Currently, CodePTMs use the tokenization and embedding methods developed in NLP. For example, they use SentencePiece as the tokenizer as well as token and position embeddings. However, code is not exactly the same as NL: code 
contains different types of lexical tokens such as variables, control symbols, and keywords. We speculate that NLP tokenization and embedding methods would not yield optimal performances for CodePTMs, and recommend that researchers look into the possibility of developing code-specific versions of these methods.


\vspace{-1mm}
\paragraph{Pre-training methods.}
Pre-training tasks that can better exploit code-specific characteristics (e.g., code structure, the presence of branches, and 
the use of different identifiers taken from a largely unrestricted vocabulary to express the same meaning) may be needed in order to train more powerful CodePTMs. Most of the existing SE-specific pre-training tasks (see Section~\ref{section:pre_training_task}) still do not completely step outside the NLP mindset. IMLM, for example, is just a version of MLM that masks identifiers, and in fact, pre-training on IMLM has even yielded worse results than pre-training on MLM for DOBF~\cite{dobf}. We believe that the design of code-specific pre-training methods is currently limited in part by the NLP tokenization and embedding methods that are currently in use, and that a fundamental overhaul in the design of code-specific pre-training methods that involves designing code-specific tokenization and embedding methods will likely be needed. 

\subsection{Learning Code Form and Functionality}
\label{section:outlook_roles_code_input}

Code has both {\em form}, which is defined by combinations of particular code identifiers, and {\em function}, which is independent of any particular code identifiers~\cite{contracode}. Note that the CodePTMs listed in Table~\ref{table:codeptm_detail} all learn representations of source code from the ``form'' instead of the ``function'' perspective. Learning code functionality, however, will undoubtedly help CodePTMs understand the code better and achieve higher performances on SE tasks. So we believe that designing CodePTMs that can learn both code form and code functionality would be a valuable research direction. 

\subsection{Adaptation to Downstream Tasks}
\label{section:outlook_adaptation}

Currently, fine-tuning is the primary method for transferring the knowledge acquired during pre-training to downstream tasks. However, fine-tuning can be inefficient because all model parameters need to be updated. To mitigate this problem, the NLP community has proposed several solutions, such as (1) model reprogramming (i.e., freezing the original parameters of the PTMs and adding small fine-tunable adaption modules for specific tasks~\cite{chen2022model}), (2) using prompt tuning~\cite{brown2020language}, and (3) using model compression (e.g., pruning and knowledge distillation). How to adapt or extend these methods for fine-tuning CodePTMs is a promising research direction.

\subsection{CodePTMs for Niche Applications}
\label{section:outlook_nich_applications}

Rather than attempting to design a single CodePTM that works well on all SE tasks, we recommend that specialized CodePTMs be designed for different classes of SE tasks (e.g., Understanding vs.\ Generation). Our recommendation is based on our earlier observations that different model design choices may be better suited for different kinds of tasks. For instance, theoretically speaking, TE-based models tend to work better than TD- and TF-based models on Understanding tasks, whereas the reverse is generally true for Generation tasks. One may even go as far as designing task-specific CodePTMs. The reason is that having pre-training tasks that are more similar to the downstream task at hand could enable a more effective transfer of the knowledge acquired during pre-training, as discussed previously. We believe that specialized CodePTMs have an additional advantage: they tend to be smaller and hence may be more efficient, potentially allowing us to address the efficiency issues associated with model architecture (Section~\ref{section:outlook_outside_nlp}) and fine-tuning (Section~\ref{section:outlook_adaptation}).

\subsection{Unified Evaluation and Analysis}
\label{section:outlook_leaderboard}

Our understanding of the strengths and weaknesses of existing CodePTMs is currently limited by the tasks on which they are evaluated. To better understand CodePTMs, it is important to conduct a systematic evaluation of all CodePTMs on all the benchmark datasets associated with the 18 SE tasks we discussed. In addition to a comprehensive quantitative evaluation, a qualitative analysis that involves analyzing the common errors made by each model would be important.


\section*{Acknowledgments}

We thank the three anonymous reviewers for their helpful comments on an earlier draft of this paper.
This work was supported in part by the National Natural Science Foundation of China (No.\ 61802167) and the US National Science Foundation (Grant IIS-1528037). Any opinions, findings, conclusions or recommendations expressed in this paper are those of the authors and do not necessarily reflect the views or official policies, either expressed or implied, of the funding agencies. Chuanyi Li is the corresponding author.

\bibliographystyle{named}
\bibliography{ijcai22}

\end{document}

%% file: table_tasks.tex
\begin{table*}[t!]
    \centering
    \resizebox{\textwidth}{!}{
    \begin{tabular}{lp{0.9cm}lp{10.5cm}lp{2.6cm}}
        \toprule
            \textbf{Type} & 
            \textbf{I-O} & 
            \textbf{Task} & 
            \textbf{Definition} &
            \textbf{ID - Dataset} &
            \textbf{Metrics} \\
        \midrule
            \multirow{17}{*}{Und.} &
            \multirow{10}{*}{C-V} & 
            \multirow{2}{*}{WB} & 
            \textbf{\underline{W}rong \underline{B}inary Operator}: Check if a given piece of code contains any incorrect binary operators. & 
            \multirow{2}{*}{K1 - Kanade et al.~\shortcite{cubert}} & 
            \multirow{2}{*}{Acc} \\ 
        \cline{3-6}
            & 
            & 
            ET & 
            \textbf{\underline{E}xception \underline{T}ype}: Predict the precise exception type. & 
            K1 - Kanade et al.~\shortcite{cubert} & 
            Acc \\
        \cline{3-6}
            &
            & 
            \multirow{2}{*}{BD} & 
            \multirow{2}{*}{\tabincell{l}{\textbf{\underline{B}ug \underline{D}etection / Defect Detection}: Check if a given function contains a \\defect.}} & 
            D1 - Devign~\shortcite{zhou2019devign} & 
            Acc \\
        \cline{5-6}
            &
            & 
            &   
            & 
            P1 - Pradel et al.~\shortcite{pradel2018deepbugs} & 
            Acc \\ 
        \cline{3-6}
            &
            & 
            \multirow{2}{*}{CD} & 
            \multirow{2}{*}{\tabincell{l}{\textbf{\underline{C}lone \underline{D}etection}: Determine whether two code snippets are semantically\\ equivalent.}} & 
            B1 - BigCloneBench~\shortcite{svajlenko2014towards} & 
            F1 \\ 
        \cline{5-6}
            &
            &    
            &   
            & 
            C1 - CLCDSA~\shortcite{nafi2019clcdsa} & 
            P/R/F1 \\
        \cline{3-6}
            &
            & 
            CC & 
            \textbf{\underline{C}ode \underline{C}lassification}: Classify the category of a given function. & 
            P2 - POJ-104~\shortcite{mou2016convolutional} & 
            Acc/MAP@R \\
        \cline{3-6}
            &
            & 
            \multirow{2}{*}{FD} &
            \textbf{\underline{F}unction-\underline{D}ocstring Mismatch}: Determine whether a given function and the docstring correspond to each other. & 
            \multirow{2}{*}{K1 - Kanade et al.~\shortcite{cubert}} & 
            \multirow{2}{*}{Acc} \\ 
        \cline{2-6}
           & 
           \multirow{5}{*}{C-C} & 
           \multirow{2}{*}{CR} &
           \multirow{2}{*}{\tabincell{l}{\textbf{\underline{C}ode-to-Code \underline{R}etrieval}: Retrieve semantically similar code for a given \\piece of query code.}} & 
           C1 - CLCDSA~\shortcite{nafi2019clcdsa} & 
           Acc/MRR/NDCG \\ 
        \cline{5-6}
           &
           &    
           &   
           & 
           P2 - POJ-104~\shortcite{mou2016convolutional} & 
           MAP@R \\ 
        \cline{3-6}
            &
            & 
            \multirow{2}{*}{VM} & 
            \textbf{\underline{V}ariable-\underline{M}isuse Localization and Repair}: Identify the location of a misused variable and return the correct one. &
            \multirow{2}{*}{V1 - Vasic et al.~\shortcite{vasic2019neural}} & 
            \multirow{2}{*}{Acc} \\ 
        \cline{3-6}
           &
           & 
           CT & 
           \textbf{\underline{C}loze \underline{T}est}: Predict the masked token from code. & 
           D2 - De Sousa et al.~\shortcite{javabert} & 
           Acc \\ 
        \cline{2-6}
           &
           \multirow{2}{*}{NL-C} & 
           \multirow{2}{*}{CS} &
           \multirow{2}{*}{\begin{tabular}[c]{@{}l@{}}\textbf{\underline{C}ode \underline{S}earch / Text-to-Code Retrieval}: Find the most relevant piece of\\ code from a set of candidates for a given natural language description.\end{tabular}} &
           C2 - CodeSearchNet~\shortcite{husain2019codesearchnet} & 
           MRR \\ 
        \cline{5-6}
           &
           &  
           &   
           & 
           C3 - AdvText~\shortcite{lu2021codexglue} &
           MRR/F1/Acc \\ 
        \hline
            \multirow{17}{*}{Gen.} &
            \multirow{9}{*}{C-C} & 
            \multirow{3}{*}{CP} &
            \multirow{3}{*}{\tabincell{l}{\textbf{\underline{C}ode Com\underline{p}letion}: Predict the missing/following token(s) of a given code\\ context.}} & 
            S1 - Svyatkovskiy et al.~\shortcite{gpt-c} & 
            RL/EditSim. \\ 
        \cline{5-6}
            &
            &    
            &   
            & 
            L1 - Liu et al.~\shortcite{cuglm} & 
            Acc \\ 
        \cline{5-6}
            &
            &    
            &      
            & 
            A1 - Alon et al.~\shortcite{alon2020structural} & 
            Acc@k \\ 
        \cline{3-6}
            &
            & 
            \multirow{3}{*}{TL} & 
            \multirow{3}{*}{\tabincell{l}{\textbf{Code \underline{T}rans\underline{l}ation}: Translate the code in one programming language to \\the code in another programming language.}} & 
            C4 - Chen et al.~\shortcite{chen2018tree} & 
            BLEU/Acc/CBLEU \\
        \cline{5-6}
            &
            &    
            &     
            & 
            T1 - TransCorder~\shortcite{roziere2020unsupervised} & 
            Acc \\ 
        \cline{5-6}
            &
            &     
            &       
            & 
            C1 - CLCDSA~\shortcite{nafi2019clcdsa} & 
            BLEU/RL/CIDER \\ 
        \cline{3-6}
            &
            & 
            BF & 
            \textbf{\underline{B}ug \underline{F}ixing}: Repair buggy code by generating the correct version. & 
            T2 - Tufano et al.~\shortcite{tufano2019empirical} & 
            BLEU/Acc/CBLEU \\ 
        \cline{3-6}
            &
            & 
            MG & 
            \textbf{\underline{M}utant \underline{G}eneration}: Inject in working code a mutant for a real bug. & 
            T3 - Tufano et al.~\shortcite{tufano2019learning} & 
            Acc \\ 
        \cline{3-6}
            &
            & 
            AG & 
            \textbf{\underline{A}ssert \underline{G}eneration}: Generate a correct unit test assert statement. & 
            W1 - Watson et al.~\shortcite{watson2020learning} & 
            Acc@k \\ 
        \cline{2-6}
            &
            \multirow{7}{*}{C-NL} & 
            \multirow{5}{*}{SU} &
            \multirow{5}{*}{\tabincell{l}{\textbf{Code \underline{Su}mmarization / Code Documentation}: Generate a textual descrip-\\tion that describes the functionality of a function.}} & 
            C2 - CodeSearchNet~\shortcite{husain2019codesearchnet} & 
            BLEU \\ 
        \cline{5-6}
            &
            &    
            &    
            & 
            H1 - Haque et al.~\shortcite{haque2020improved} & 
            BLEU/RL \\ 
        \cline{5-6}
            &
            &    
            &             
            & 
            H2 - Hu et al.~\shortcite{hu2018deep} & 
            BLEU \\ 
        \cline{5-6}
            &
            &    
            &         
            & 
            H3 - Hu et al.~\shortcite{hu2018summarizing} & 
            BLEU/METEOR \\ 
        \cline{5-6}
            &
            &    
            &          
            & 
            M1 - Miceli et al.~\shortcite{miceli2017parallel} & 
            BLEU \\ 
        \cline{3-6}
            &
            & 
            \multirow{2}{*}{MN} &
            \multirow{2}{*}{\tabincell{l}{\textbf{\underline{M}ethod \underline{N}aming / Extreme Code Summarization}: Predict the function \\name of a given function body.}} & 
            A2 - Allamanis et al.~\shortcite{allamanis2016convolutional} & 
            P/R/F1 \\ 
        \cline{5-6}
            &
            &  
            &          
            & 
            E1 - ETH Py150~\shortcite{raychev2016probabilistic} & 
            P/R/F1 \\ 
        \cline{2-6}
            & 
            NL-C & 
            CG &
            \textbf{\underline{C}ode \underline{G}eneration}: Generate code given a natural language description. & 
            C5 - CONCODE~\shortcite{iyer2018mapping} & 
            BLEU/Acc/CBLEU \\
        \bottomrule
    \end{tabular}
    }
    \caption{Categorization of the 18 SE tasks to which CodePTMs have been applied.}
    \label{table:application}
\end{table*}

%% file: table_pre_training_task.tex
\begin{table*}[t!]
    \centering
    \resizebox{\textwidth}{!}{
    \begin{tabular}{l|l|l|ll}
        \toprule
            \textbf{} & \multicolumn{2}{c|}{\textbf{Type}} & \textbf{Task} & \textbf{Full Name and Description} \\
        \midrule
            \multirow{10}{*}{\tabincell{l}{N\\L\\P}}  & \multicolumn{2}{c|}{\multirow{3}{*}{LM}}  & FLM ~\shortcite{gpt-c}       
            & Forward LM: maximizes the conditional probabilities of all the words by taking their previous words as contexts. \\ \cline{4-5}
                                   & \multicolumn{2}{c|}{}                     & FNP~\shortcite{prophetnet-code}       
            & Future N-gram Prediction: a variant of FLM that involves predicting the next $n$ ($n>1$) tokens simultaneously instead of one token. \\ \cline{4-5}
                                  & \multicolumn{2}{c|}{}                     & BiLM  ~\shortcite{scelmo}     
            & Bidirectional LM: combines a forward LM and a backward LM, and jointly maximizes the likelihood of the tokens both directions. \\ \cline{2-5}           
                                  & \multicolumn{2}{c|}{\multirow{4}{*}{MLM}} & BMLM ~\shortcite{javabert}  
            & Basic version of MLM: randomly masks a certain percentage of tokens in the input, then predicts the masked tokens.\\ \cline{4-5}
                                  & \multicolumn{2}{c|}{}                     & WWM ~\shortcite{c-bert}
            & Whole Word Masking: if a word is masked, mask all subwords/tokens in it; then predict these masked tokens.\\ \cline{4-5}
                                  & \multicolumn{2}{c|}{}                     & MASS ~\shortcite{spt-code}      
            & MAsked Seq2Seq: reconstructs the sentence fragment given the remaining part of the sentence in the encoder-decoder framework.\\ \cline{2-5}
                                 & \multicolumn{2}{c|}{}                     & SMLM ~\shortcite{t5-learning}      
            & Seq2Seq MLM: randomly masks a set of token spans in the input and sequentially predicts them in the encoder-decoder framework.\\ \cline{2-5}
                                  & \multicolumn{2}{c|}{DAE}                  & DAE  ~\shortcite{plbart}       
            & Denoising Auto-Encoding: corrupts the input (by masking, deleting tokens, etc.) and uses the model to recover the original input. \\ \cline{2-5}
                                  & \multicolumn{2}{c|}{\multirow{2}{*}{CTL}} & NSP ~\shortcite{cubert}     
            & Next Sentence Prediction: determines whether two given sentences (i.e., logical lines of code) are coherent.\\ \cline{4-5}
                                  & \multicolumn{2}{c|}{}                     & RTD ~\shortcite{codebert}   
            & Replaced Token Detection: identifies the replaced tokens in the input (i.e., tokens produced by a small generator network). \\ \cline{1-5}
            
            \multirow{16}{*}{\tabincell{l}{S\\E}}  & \multicolumn{2}{c|}{\multirow{4}{*}{CA}}  & IMLM ~\shortcite{cuglm}      
            & Identifier MLM: an adaptation of MLM to source code that masks only the identifiers in the code text.  \\ \cline{4-5}
                                & \multicolumn{2}{c|}{}                     & SIMLM  ~\shortcite{codet5}  
            & Seq2Seq IMLM: an adaptation of Seq2Seq MLM to source code that masks only the identifiers in the code text.\\ \cline{4-5}
                                  & \multicolumn{2}{c|}{}                     & IT  ~\shortcite{codet5}  
            & Identifier Tagging: determines if the input token at each position is an identifier or not via binary classification.\\ \cline{4-5}
                                  & \multicolumn{2}{c|}{}                     & CCL ~\shortcite{oscar}       
            & Code Contrastive Learning: minimizes/maximizes the distances between the representations of similar/dissimilar code snippets.\\ \cline{2-5}
                                  & \multicolumn{2}{c|}{\multirow{2}{*}{SA}}  & EP ~\shortcite{graphcodebert}     
            & Edge Prediction: masks the edges connecting randomly selected nodes in a DFG, then predicts the masked edges.  \\ \cline{4-5}
                                  & \multicolumn{2}{c|}{}                     & NOP ~\shortcite{treebert}      
            & Node Order Prediction: randomly changes the order of some nodes in an AST, then determines if a change occurs.\\ \cline{2-5}
                                  & \multirow{10}{*}{\tabincell{l}{C\\M\\A}} &\multirow{2}{*}{CN}  & BDG ~\shortcite{codet5} 
            &Bimodal Dual Generation: generates a NL summary if code is given, and generates code if NL is given. \\ \cline{4-5}
                                  &                       &                       & MNG ~\shortcite{spt-code}   
            &Method Name Generation: generates the sub-token sequence of the method name based on a given method body. \\ \cline{3-5}
                                  &                       &\multirow{7}{*}{CS}  & NA ~\shortcite{graphcodebert} 
            &Node Alignment: samples nodes in a DFG, masks the edge connecting each node to its code token, then predicts the masked edges. \\ \cline{4-5}
                                  &                       &                      & TMLM ~\shortcite{treebert} 
            &Tree MLM: masks some terminal nodes/identifiers in ASTs/code on encoder/decoder side, then generates complete code sequence. \\ \cline{4-5}
                                  &                       &                      & VGVAE ~\shortcite{codedisen} 
            & vMF-Gaussian Variational Autoencoder: disentangles code semantics from code syntax under the supervision of a masked AST.  \\ \cline{4-5}
                                  &                       &                      & CAP ~\shortcite{spt-code} 
            & Code-AST Prediction: determines whether the given code and AST correspond to each other.\\ \cline{4-5}
                                  &                       &                      & CLR  ~\shortcite{codedisen}
            &Cross-Language Reconstruction: reconstructs the code snippet in one PL from functionally equivalent code snippets in other PLs.
            \\ \cline{4-5}
                                  &                       &                      & PD ~\shortcite{codedisen} 
            & Posterior Distribution: reduces difference in distributions of functionally equiv.\ code snippets in different PLs over code semantics.
            \\ \cline{4-5}
                                  &                       &                      & ACP ~\shortcite{codedisen} 
            &Attentive Code Position: predicts the node type of a code token in an AST through an attention mechanism. \\ \cline{3-5}
                                  &                       &CNS                & MCL ~\shortcite{syncobert} 
            &Multi-modal Contrastive Learning: maximizes/minimizes the representation similarity between positive/negative samples. \\ 
        \bottomrule
    \end{tabular}
    }
    \caption{Categorization and description of the pre-training tasks used by existing CodePTMs.}
    \label{table:pretrainingtask}
   \vspace{-2mm}
\end{table*}

%% file: table_cptms_taxonomy.tex
\begin{table*}[t!]
    \centering
    \resizebox{\textwidth}{!}{
    %
    \begin{tabular}{p{9mm}|p{8mm}|p{4mm}p{3mm}p{3mm}p{7mm}|l|lp{3mm}p{3mm}p{3mm}p{3mm}p{3mm}p{3mm}p{3mm}p{3mm}p{3mm}p{3mm}p{3mm}p{3mm}p{3mm}p{3mm}p{3mm}p{3mm}p{3mm}p{3mm}} %
        \toprule
            \multirow{2}{*}{\textbf{Arch.}}
            & \multirow{2}{*}{\textbf{Mod.}}
            & \multicolumn{4}{c|}{\textbf{Pre-Training Tasks}}
            & \multirow{2}{*}{\textbf{PL}}
            & \multirow{2}{*}{\textbf{CodePTM}}
            & \multicolumn{10}{c}{\textbf{SE Understanding Tasks}}
            & \multicolumn{8}{|c}{\textbf{SE Generation Tasks}}
            \\ \cline{3-6} \cline{9-26}
            &&\textbf{NLP}&\textbf{CA}&\textbf{SA}&\textbf{CMA}&&
            &\textbf{WB}
            &\textbf{ET}
            &\textbf{BD}
            &\textbf{CD}
            &\textbf{CC}
            &\textbf{FD}
            &\textbf{CR}
            &\textbf{VM}
            &\textbf{CT}
            &\textbf{CS}
            &\multicolumn{1}{|c}{\textbf{CP}}
            &\textbf{TL}
            &\textbf{BF}
            &\textbf{MG}
            &\textbf{AG}
            &\textbf{SU}
            &\textbf{MN}
            &\textbf{CG}\\
        \midrule
            \multirow{2}{*}{LSTM}   & Uni & \checkmark & & & & Mono  & SCELMo 
            &&&\textbf{P1}&&&&&&&       &&&&&&&& \\ \cline{2-26}
            
                                    & Bi & &&&\checkmark  & Multi  & CodeDisen 
            &&&&\textbf{C1}&&&\textbf{C1}&&&             &&\textbf{C1}&&&&&&\\ \hline
                                    
            \multirow{9}{*}{TE} & \multirow{4}{*}{Uni}& \multirow{3}{*}{\checkmark} &&&  & \multirow{3}{*}{Mono}  & CuBERT 
            &\textbf{K1}&\textbf{K1}&&&&\textbf{P2}&&\textbf{V1}&&          &&&&&&&&\\ \cline{8-26}
            
                                    & &  &&&   & & C-BERT
            &&&D1&&&&&&&      &&&&&&&& \\ \cline{8-26}
                                    
                                    & & &&&  & & JavaBERT
            &&&&&&&&&\textbf{D2}&         &&&&&&&& \\ \cline{3-26}
                                    
                                    &  & \checkmark &\checkmark&&  & Multi & CugLM
            &&&&&&&&&&        &\textbf{L1}&&&&&&& \\ \cline{2-26}
                                    
                                    & \multirow{2}{*}{Bi}  & \checkmark &&&  & Multi  & CodeBERT
            &&&&&&&&&& C2            &&&&&&C2&& \\ \cline{3-26}
                                    
                                    &&\checkmark&\checkmark&& & Mono & OSCAR 
            &&&&&\textbf{P2}&&P2&&&         &&&&&&&&\\ \cline{2-26}
                                    
                                    & \multirow{2}{*}{Multi} & \checkmark&&&\checkmark  & Multi & GraphCodeBERT
            &&&&B1&&&&&& C2        &&C4&T2&&&&& \\ \cline{3-26}
                                    
                                    &  & \checkmark&\checkmark&\checkmark&\checkmark  & Multi  & SynCoBERT 
            &&&D1&\textbf{B1}&&&\textbf{P2}&&\multicolumn{3}{c}{C2,C3}&C4&&&&&&\\ \hline
                                    
            TD & Uni  & \checkmark&&&  & Multi  & GPT-C 
            &&&&&&&&&&       &\textbf{S1}&&&&&&&\\ \hline
            
            \multirow{12}{*}{TF} & \multirow{2}{*}{Uni} & &\checkmark&&  & Multi  & DOBF
            &&&&B1&&&&&& C3      &&\textbf{T1}&&&&C2&& \\ \cline{3-26}
            
                                    &  & \checkmark&&&  & Mono  & DeepDebug 
            &&&&&&&&&&           &&&T2&&&&& \\ \cline{2-26}
                                    
                                    & \multirow{6}{*}{Bi} & \multirow{4}{*}{\checkmark} &&&  & Mono & T5-learning 
            &&&&&&&&&&          &&&T2&\textbf{T3}&\textbf{W1}&\textbf{H1}&&\\ \cline{7-26}
                                    
                                    &  & &&& & \multirow{3}{*}{Multi} & PLBART 
            &&&D1&B1&&&&&&          &&C4&&&&C2&&C5\\ \cline{8-26}
                                    
                                    &  & &&&  &  & CoTexT 
            &&&D1&&&&&&&              &&&T2&&&C2&&C5\\ \cline{8-26}
                                    
                                    &   & &&&  &  & ProphetNet-Code 
            &&&&&&&&&&                 &&&&&&\textbf{C2}&&\\ \cline{3-26}
                                    
                                    &   & \checkmark&\checkmark&&\checkmark  & Multi  & CodeT5 
            &&&\textbf{D1}&B1&&&&&&                 &&\textbf{C4}&\textbf{T2}&&&\textbf{C2}&&\textbf{C5}\\ \cline{3-26}
                                    
                                    &   & &&\checkmark&\checkmark & Multi  & TreeBERT 
            &&&&&&&&&&             &&&&&&\textbf{H2}&\multicolumn{2}{l}{\textbf{A2},\textbf{E1}}\\ \cline{2-26}
                                    
                                    & Multi & \checkmark&&&\checkmark  & Multi  & SPT-Code 
            &&&&&&&&&& C2             &\textbf{A1}&\textbf{C4}&T2&&\multicolumn{3}{c}{C2,\textbf{H3},\textbf{M1}}&\\
        \bottomrule
    \end{tabular}
    }
    \caption{Categorization of existing CodePTMs along four dimensions and their performances on downstream SE tasks. If a CodePTM is applied to a task, we list the ID of the benchmark dataset on which the CodePTM was evaluated (see Table~\ref{table:application} for the ID associated with each dataset), boldfacing the ID if the CodePTM achieved SOTA results on the corresponding dataset.}
    \label{table:taxonomy}
    \vspace{-2mm}
\end{table*}

%% file: table_cptms_details.tex
\begin{table*}[t!]
    \centering
    \resizebox{\textwidth}{!}{
    \begin{tabular}{p{4cm}p{4cm}lp{5.1cm}l}
        \toprule
            \textbf{CodePTM}
            & \textbf{Input}
            & \textbf{Objective}
            & \textbf{Dataset}
            & \textbf{Dataset Size} \\
        \midrule
            SCELMo~\shortcite{scelmo}                   & Code                   & BiLM                                  & JS GitHub Repos           & 150K Files  \\ \hline
            CodeDisen~\shortcite{codedisen}             & Code + AST Seq         & VGVAE + CLR + PD + ACP                      & CLCDSA                              & 26K Functions  \\ \hline
            CuBERT~\shortcite{cubert}                   & Code                   & BMLM + NSP                             & Python from BigQuery                      & 7.4M Files \\ \hline
            C-BERT~\shortcite{c-bert}                   & Code                   & WWM                                   & C GiHub Repos                              & 5.8GB    \\ \hline
            JavaBERT~\shortcite{javabert}               & Code                   & BMLM                                   & Java GitHub Repos                         & 3M Files    \\ \hline
            CugLM~\shortcite{cuglm}                     & Code                   & IMLM + NSP + FLM                   & Java, TS GitHub Repos                       & 617K Files \\ \hline
            CodeBERT~\shortcite{codebert}               & Code + Doc             & BMLM \& RTD                            & CodeSearchNet                            & 6.5M Functions \\ \hline
            OSCAR~\shortcite{oscar}                     & IR + AEI               & BMLM + CCL                             & C/C++ GitHub Repos                         & 500K Functions  \\ \hline
            GraphCodeBERT~\shortcite{graphcodebert}     & Code + Doc + DFG Nodes & BMLM + EP + NA                             & CodeSearchNet (Bimodal)                     & 2.3M Functions\\ \hline
            SynCoBERT~\shortcite{syncobert}             & Code + Doc + AST Seq   & BMLM + IT + TEP + MCL                            & CodeSearchNet                              & 6.5M Functions      \\ \hline
            GPT-C~\shortcite{gpt-c}                     & Code                   & FLM                            & Python, C\#, JS/TS GitHub Repos            & 4.7M Files    \\ \hline
            DOBF~\shortcite{dobf}                       & Code                   & SIMLM(Seq2Seq IMLM)                         & Java, Python from BigQuery               & 11.5M Files    \\ \hline
            DeepDebug~\shortcite{deepdebug}             & Code                   & SMLM(Seq2Seq MLM)                           & Java GitHub Repos                          & 8M Files   \\ \hline
            T5-learning~\shortcite{t5-learning}         & Code                   & SMLM(Seq2Seq MLM)                           & CodeSearchNet (Java)                  & 1.5M Functions          \\ \hline
            \multirow{2}{*}{PLBART~\shortcite{plbart}}  & \multirow{2}{*}{Code \& Posts}  & \multirow{2}{*}{DAE (masking / deletion / infilling)}  & Java, Python GitHub Repos  & 680M Functions\\ \cline{4-5}
                                                        &                        &                                       & StackOverflow Posts                             & 47M Posts \\ \hline
            \multirow{2}{*}{CoTexT~\shortcite{cotext}}  & \multirow{2}{*}{Code + Doc}     & \multirow{2}{*}{SMLM(Seq2Seq MLM)} & CodeSearchNet                               & 6.5M Functions \\ \cline{4-5}
                                                        &                        &                                       & Java, Python from BigQuery                  & 6.4M Functions \\ \hline
            ProphetNet-Code~\shortcite{prophetnet-code} & Code \& Doc            & FNP                                   & CodeSearchNet (Bimodal)                      & 2.3M Functions        \\ \hline
            \multirow{2}{*}{CodeT5~\shortcite{codet5}}  & \multirow{2}{*}{Code + Doc} & SMLM(Seq2Seq MLM) / IT /                  & CodeSearchNet                               & 6.5M Functions      \\ \cline{4-5}
                                                        &                        & SIMLM(Seq2seq IMLM) / BDG                      & C, C\# from BigQuery                        & 1.85M Functions      \\ \hline
            TreeBERT~\shortcite{treebert}               & Code + AST Paths       & TMLM + NOP                              & Java, Python from BigQuery                  & 21.3M Files      \\ \hline
            SPT-Code~\shortcite{spt-code}               & Code + Names + AST Seq & CAP \& MASS \& MNG                        & CodeSearchNet                                 & 6.5M Functions              \\ 
        \bottomrule
    \end{tabular}
    }
    \caption{Details of how the CodePTMs are pre-trained. The pre-training scheme employed for each CodePTM is characterized by (1) the input modalities (if multiple modalities are involved, they can be handled via a Together (+) or Standalone (\&) strategy); (2) the pre-training objectives (if multiple pre-training
    objectives are involved, they can be learned jointly (+), sequentially (\&), or alternately (/)); (3) the dataset on which the CodePTM is pre-trained; and (4) the size of the dataset.
    }
    \vspace{-2mm}
    \label{table:codeptm_detail}
\end{table*}

%% file: table_sota_new.tex
\begin{table}[t!]
    \centering
    \resizebox{\linewidth}{!}{
    \begin{tabular}{p{0.4cm}p{0.4cm}lll}
        \toprule
            \textbf{Task} & 
            \textbf{DS} &
            \textbf{Best CodePTM}&
            \textbf{Best non-CodePTM} &
            \textbf{ER}\\
        \midrule
            WB & 
            K1& 
            82.3 (CuBERT~\shortcite{cubert})& 
            73.8 (GREAT~\shortcite{hellendoorn2020global}) &
            32.4
            \\
        \hline 
            ET & 
            K1& 
            79.1 (CuBERT~\shortcite{cubert})& 
            49.5 (Transformer~\shortcite{cubert}) &
            58.6
            \\
        \hline 
            BD & 
            D1 & 
            65.7 (CodeT5~\shortcite{codet5})& 
            62.4 (code2vec~\shortcite{coimbra2021using}) &
            8.8
            \\
        \hline 
            \multirow{2}{*}{CD} & 
            B1& 
            97.4 (SynCoBERT \shortcite{syncobert})& 
            95.0 (FA-AST~\shortcite{wang2020detecting}) & 
            48.0
            \\ 
        \cline{2-5} 
            & 
            C1 & 
            90.0 (CodeDisen~\shortcite{codedisen}) & 
            81.0 (Tree-LSTM~\shortcite{shido2019automatic} ) & 
            47.3
            \\ 
        \hline 
            CC & 
            P2 & 
            98.0 (OSCAR~\shortcite{oscar}) & 
            96.6 (ProGraML~\shortcite{oscar}) & 
            43.2
            \\
        \hline 
            FD &
            K1& 
            98.0 (CuBERT~\shortcite{cubert}) & 
            91.0 (Transformer~\shortcite{cubert}) &
            78.7
            \\ 
        \hline 
            \multirow{2}{*}{CR} &
            C1& 
            31.6 (CodeDisen~\shortcite{codedisen})& 
            16.6 (Pontes et al.~\shortcite{pontes2018predicting} ) &
            17.9
            \\ 
        \cline{2-5} 
            & 
            P2&
            88.2 (SynCoBERT~\shortcite{syncobert})& 
            82.4 (MISIM~\shortcite{ye2020misim} ) &
            32.9
            \\ 
        \hline 
            VM & 
            V1 & 
            95.2 (CuBERT~\shortcite{cubert}) & 
            80.5 (BiLSTM~\shortcite{cubert}) &
            75.4
            \\ 
        \hline 
            CT & 
            D2 & 
            94.4 (JavaBERT~\shortcite{javabert})& 
            $-$&
            
            \\
        \hline 
            \multirow{2}{*}{CS} &
            C2 & 
            74.0 (SynCoBERT~\shortcite{syncobert})& 
            41.9 (Transformer~\shortcite{graphcodebert}) &
            55.2
            \\ 
        \cline{2-5}   
            & 
            C3 & 
            38.1 (SynCoBERT~\shortcite{syncobert}) &
            $-$ &
            
            \\
        \hline 
            \multirow{3}{*}{CP} &
            S1 & 
            82.8 (GPT-C~\shortcite{gpt-c}) & 
            $-$&
            
            \\ 
        \cline{2-5} 
            & 
            L1 & 
            81.9 (CugLM~\shortcite{cuglm}) & 
            71.7 (Transformer-XL~\shortcite{dai2019transformer}) &
            36.0
            \\ 
        \cline{2-5}       
            & 
            A1 & 
            26.5 (SPT-Code~\shortcite{spt-code})& 
            24.7 (SLM~\shortcite{alon2020structural}) &
            2.4
            \\ 
        \hline 
            \multirow{3}{*}{TL} & 
            C4 & 
            66.4 (CodeT5~\shortcite{codet5}) & 
            35.4 (Transformer~\shortcite{plbart}) &
            47.9
            \\
        \cline{2-5} 
            & 
            T1& 
            41.8 (DOBF~\shortcite{dobf}) & 
            34.7 (Transformer~\shortcite{dobf}) &
            10.9
            \\ 
        \cline{2-5}      
            & 
            C1 & 
            29.7 (CodeDisen~\shortcite{codedisen}) & 
            25.8 (Tree-LSTM~\shortcite{shido2019automatic}) &
            5.3
            \\ 
        \hline 
            BF & 
            T2 & 
            18.3 (CodeT5~\shortcite{codet5}) & 
            12.7 (S2S+COPY~\shortcite{panthaplackel2021copy}) &
            6.5
            \\ 
        \hline 
            MG & 
            T3 & 
            28.0 (T5-learning~\shortcite{t5-learning}) & 
            17.0 (Tufano~\shortcite{tufano2019learning}) &
            13.3
            \\ 
        \hline 
            AG & 
            W1 & 
            66.0 (T5-learning~\shortcite{t5-learning}) & 
            65.0 (Watson et al.~\shortcite{watson2020learning}) &
            2.9
            \\ 
        \hline 
            \multirow{5}{*}{SU} &
            C2 & 
            19.7 (CodeT5~\shortcite{codet5}) & 
            15.5 (Transformer~\shortcite{cubert}) &
            4.9
            \\ 
        \cline{2-5}    
            & 
            H1 & 
            21.0 (T5-learning~\shortcite{t5-learning})& 
            19.0 (Haque et al.~\shortcite{haque2020improved}) &
            2.5
            \\ 
        \cline{2-5}
            & 
            H2 & 
            20.4 (TreeBERT~\shortcite{treebert})& 
            19.7 (GNN+GRU~\shortcite{leclair2020improved}) &
            0.9
            \\ 
        \cline{2-5}      
            & 
            H3 & 
            49.1 (SPT-Code~\shortcite{spt-code})& 
            48.2 (AST-Trans~\shortcite{tang2022ast}) &
            1.6
            \\ 
        \cline{2-5}    
            & 
            M1 & 
            36.1 (SPT-Code~\shortcite{spt-code})& 
            34.7 (AST-Trans~\shortcite{tang2022ast}) &
            2.1
            \\ 
        \hline  
            \multirow{2}{*}{MN} &
            A2& 
            60.1 (TreeBERT~\shortcite{treebert}) & 
            57.5 (GNN+GRU~\shortcite{leclair2020improved}) &
            6.1
            \\ 
        \cline{2-5}       
            & 
            E1 & 
            39.0 (TreeBERT~\shortcite{treebert}) & 
            34.4 (GNN+GRU~\shortcite{leclair2020improved}) &
            7.0
            \\ 
        \hline 
            CG &
            C5 & 
            22.3 (CodeT5~\shortcite{codet5}) & 
            12.2 (Iyer et al.~\shortcite{iyer2019learning}) &
            11.5
            \\
        \bottomrule
    \end{tabular}
    }
    \caption{Relative error reduction rates achieved by CodePTMs on SE tasks/datasets. "DS" shows the commonly used evaluation datasets for each SE task (see Table~\ref{table:application} for details on these datasets). "Best CodePTM" shows the best result achieved to date by a CodePTM on the corresponding dataset and the name of the CodePTM. "Best non-CodePTM" shows the best result achieved to date by an approach that does not involve pre-training on the corresponding dataset and the name of the approach (note that ``\textbf{$-$}'' indicates that non-CodePTM-based approaches have not been applied to the corresponding dataset). "ER" shows the relative error reduction rate for each dataset. Information on the evaluation metric used for each dataset can be found in Table~\ref{table:application}.}
    \label{table:sota}
    \vspace{-2mm}
\end{table}